\documentstyle[preprint,aps]{revtex}
\begin{document}
\newcommand{\beq}{\begin{equation}}
\newcommand{\eeq}{\end{equation}}
\newcommand{\xplus}{x^{+}}
\newcommand{\xminus}{x^{-}}
\newcommand{\half}{{1\over 2}}
\newcommand{\fourth}{{1\over 4}}
\newcommand{\emath}{\mathrm{e}}
\newcommand{\albar}{\bar{\alpha}}
\newcommand{\bbar}{\bar{b}}
\newcommand{\Jbar}{\bar{J}}
\newcommand{\Omegabar}{\bar{\Omega}}
\newcommand{\pbar}{\bar{p}}
\newcommand{\phibar}{\bar{\phi}}
\newcommand{\Phibar}{\bar{\Phi}}
\newcommand{\psibar}{\bar{\psi}}
\newcommand{\qbar}{\bar{q}}
\newcommand{\wbar}{\bar{w}}
\newcommand{\xibar}{\bar{\xi}}
\newcommand{\zbar}{\bar{z}}
\newcommand{\hO}{\hat{O}}
\newcommand{\hU}{\hat{U}}
\newcommand{\cA}{{\mathcal{A}}}
\newcommand{\cH}{{\mathcal{H}}}
\newcommand{\cL}{{\mathcal{L}}}
\newcommand{\cO}{{\mathcal{O}}}
\newcommand{\cP}{{\mathcal{P}}}
\newcommand{\lrarrow}[1]{\raisebox{1.4ex}{\makebox[0pt][l]
{$\leftrightarrow$}}#1}
\newcommand{\eqref}[1]{(\ref{#1})}

\draft
\preprint{\vbox{\hbox{KUNS-1433}\hbox{HE(TH) 97/01}}}
\title{Smooth Bosonization as a Quantum Canonical Transformation}
\author{Andrew J. Bordner\thanks{e-mail address: 
bordner@gauge.scphys.kyoto-u.ac.jp} \\
Department of Physics, Kyoto University, Kyoto 606-01, Japan}
\date{\today}
\maketitle
\begin{abstract}
We consider a $1+1$ dimensional field theory which contains both a
complex fermion field and a real scalar field.  
We then construct a
unitary operator that, by a similarity transformation, gives a
continuum of equivalent theories which smoothly interpolate between 
the massive Thirring model and the sine-Gordon model.
This provides an implementation of smooth
bosonization proposed by Damgaard {\it et al.} as well as an example 
of a quantum
canonical transformation for a quantum field theory.  
\end{abstract}
\section{Introduction}
Duality, or the quantum equivalence of field theories, allows one to
relate quantities in one theory, such as the particle spectrum and Green's 
functions, to those of another theory.  This concept is most useful
when duality maps a theory with a strong coupling, in which
perturbation theory is invalid, to one with a weak coupling, in which
a perturbative calculation may be performed.
Unfortunately, most such duality transformations are hypothetical
since an explicit operator mapping is absent.  Duality is usually
demonstrated either by symmetry arguments, as in non-Abelian
bosonization \cite{NA_bosonization}, by a path integral calculation, 
as in S- duality \cite{S_duality} and Abelian bosonization
\cite{path_int_boson}, or by a linear
canonical transformation, as in T-duality \cite{T_duality}.  

The equivalence of two different quantum mechanical theories may be
rigorously established by means of a similarity transformation of all quantum
operators.  These transformations were established by Dirac as the
quantum version of canonical transformations in classical mechanics;
the former preserving the quantum commutators and the latter the
Poisson brackets.  More recently, Anderson further investigated these
mappings, which he named ``quantum canonical transformations'', and
emphasized that they need not be unitary as originally
presumed but rather isometric \cite{Anderson}.  Quantum canonical 
transformations map
operators $\hO_{i}$ in one theory to $\hO'_{i}$ in another quantum
theory by $\hO_{i}'= \hU \hO_{i} \hU^{-1}$ with $\hU$ an operator
composed of products of the canonical position and momentum
operators.  Such a transformation obviously preserves commutators of
operators and both theories have the same energy spectrum with eigenstates
mapped as $|E\rangle'= \hU |E\rangle$.

In this letter, we propose to extend quantum canonical transformations 
to $1+1$ dimensional quantum field theories, namely the Abelian 
bosonization of free
bosons and fermions as well as the equivalence of the massive Thirring 
model and the sine-Gordon model.  Abelian bosonization is an ideal
test of this technique since the operator correspondences between the
fermionic and bosonic theories are known
\cite{Coleman,Mandel_Klaiber}.  We will consider a theory which
contains both fermions and bosons.  A quantum canonical transformation 
of a theory with only fermions, the massive
Thirring model, yields a
continuum of theories with interacting fermions and bosons as well as
the corresponding theory containing only bosons, the sine-Gordon model.   

\section{Quantum canonical transformations for a field theory}
We first review quantum canonical transformations in quantum
mechanics and next examine how to extend this definition to a quantum
field theory.  We begin by considering quantum mechanics in the
Schroedinger picture, with only the states evolving with time.  A
quantum canonical transformation is a similarity
transformation of the quantum operators, which may be defined by its
action on the canonical position and momentum operators $q_{i}$ and
$p_{i}$ in the original theory, to obtain $q'_{i}$ and
$p'_{i}$ in the new theory 
\beq
q'_{i}\equiv U(q,p)\:q_{i}\:U^{-1}(q,p),\qquad p'_{i}\equiv U(q,p)
\:p_{i}\:U^{-1}(q,p).
\eeq
If the states also transform as $|\psi\rangle'\equiv
U(q,p)|\psi\rangle$ then matrix elements are preserved and $U(q,p)$ is 
an isometry.  This transformation is specified by an operator
$U(q,p)$, which we
assume to be invertible.  A quantum canonical transformation also
preserves the canonical commutation relations
\beq
[q'_{i},p'_{j}]=[q_{i},p_{j}]=\imath\delta_{ij} ,
\eeq
with all other commutators vanishing.  These transformations may be used
to calculate some physical properties of a quantum mechanical system,
for instance the energy eigenstates and eigenvalues, by mapping a
given model to another model whose solutions are known, such as the
harmonic oscillator \cite{Anderson}.

Quantum field theory is usually formulated in the interaction
picture where the Hamiltonian density operator is divided into 
the sum of a free part 
$\cH_{0}$ and an interacting part $\cH_{int}$.  The free field
$\phi(x)$ satisfies the operator equation ${\partial\over 
\partial_{t}}\phi(\vec{x},t)=\imath[\int d\vec{y}\: 
\cH_{0}(\vec{y},t),\phi(\vec{x},t)]$.  
 We will assume that
exact physical quantities, {\it i.e.}, calculated to all orders in
perturbation theory, are independent of the choice of this
partition of the Hamiltonian.
The Green's functions for the interacting fields $\phi_{I}(\vec{x},t)$ may be
expressed in terms of the free fields using the well-known equation 
\begin{eqnarray}
\label{interact_Green}
G(x_{1},\ldots,x_{n})&\equiv& \langle 0 | T \left[\phi_{I}(\vec{x}_{1},t_{1})
\cdots\phi_{I}(\vec{x}_{n},t_{n}) \right] | 0 \rangle \\ \nonumber
&=&{{\langle 0 | T^{*} \left[ \phi(\vec{x}_{1},t_{1})\cdots 
\phi(\vec{x}_{n},t_{n}) 
\exp\left(-\imath\int d\vec{x}'dt' \cH_{int}(\phi(\vec{x}',t'))
\right)\right]| 0 \rangle}\over
{\langle 0 | T^{*} \exp\left(-\imath\int d\vec{x}'dt'\cH_{int}
(\phi(\vec{x}',t'))\right)| 0 \rangle}} .
\end{eqnarray}
$T^{*}$ operator products are necessary in this equation for the
theory that we will
consider, the Thirring model, because of the Schwinger term 
in the current
algebra.  However, following the arguments of Ref. \cite{Banks}, which 
notes that 
Eqn. \eqref{interact_Green} yields Lorentz covariant
expressions if the $T^{*}$ product is replaced by the naive
time-ordered product and a covariant renormalization scheme is
employed, we consider usual time-ordered products for this equation.
Thus the quantities that
we would like to transform are vacuum
expectation values of time-ordered
products of the free fields $\phi(\vec{x},t)$.
Rather than using the algebra of the canonical operators at equal time 
in the Heisenberg picture, or equivalently in the Schroedinger
picture, to construct the operator $U$ which implements the
transformation, we may use the commutator of free fields at different
times $[\phi(\vec{x}_{1},t_{1}),\phi(\vec{x}_{2},t_{2})]\equiv  
\Delta(\vec{x}_{2}-\vec{x}_{1},t_{2}-t_{1})$.  $\Delta(\vec{x},t)$ is
a c-number function.  Since it appears difficult to 
construct $U$ for a general theory we will consider 
here the simpler case of a massless free field $\phi$ in 1+1
spacetime dimensions for which the 
powerful methods of conformal field theory may be applied.

We consider the field theory defined on a circle, {\it i.e.}, periodic 
as $x^{1}\rightarrow x^{1} + \sqrt{2}L$ to avoid the infrared
divergences appearing in massless field theories.  This periodicity
also allows a bijective map of the spacetime coordinates to the
complex plane with the origin excluded.
Light cone coordinates are defined as $x^{\pm}\equiv{1\over\sqrt{2}}
(x^{0}\pm x^{1})$.  After a Wick rotation from Minkowski space to
Euclidean space $x^{0}_{E}=-\imath x^{0}$ one 
defines $z\equiv\exp\left({2\pi\imath\over L}\xplus_{E}\right)$ and
likewise defines $\zbar$ in terms of $\xminus_{E}$.  Hereafter we consider
only time-ordered products of operators, corresponding to
radial-ordered products in the conformal field
theory, since only these have well-defined
vacuum expectation values in the Euclidean theory for any value of
spacetime coordinates of the operators\cite{R_quantization}.  We then 
define a general 
form for $U$ in the
conformal field theory as $U\equiv R \exp \imath \left( \oint_{C}
{dz\over {2\pi\imath}}\Omega(z) + \oint_{C}
{d\zbar\over {2\pi\imath}}\Omegabar(\zbar)\right)\equiv R \exp 
\imath \cA$. The integration
contour $C$ is chosen so that all operator products are
radial-ordered.  If the operators
$\Omega$ and $\Omegabar$ are Hermitian then $U$ is unitary.  The
quantum canonical transformation of an operator $\cO(z,\zbar)$ is
defined as 
\begin{eqnarray}
\label{operator_trans}
\cO'(z,\zbar) &=& R \left(U^{-1}\:\cO(z,\zbar)\:U\right) \\ \nonumber
&=& R\left(\cO(z,\zbar)+\imath[\cA,\cO(z,\zbar)]
+{\imath^{2}\over 2!}[\cA,[\cA,\cO(z,\zbar)]] 
+\ldots\right) .
\end{eqnarray}
The second equality follows from the Baker-Campbell-Hausdorff
relation.  Since the two contour integrals in the commutator are
equivalent to a single contour integral about $z$ this transformation
may be easily calculated, given the Wick contractions for the relevant fields.
The resulting quantum canonical
transformation for the conformal field theory looks superficially
similar to the case for quantum mechanics in that the free fields 
$\phi(z,\zbar)$,
Hamiltonian density $\cH(\phi(z,\zbar))$ and vacuum state transform as 
\begin{eqnarray}
\phi'(z,\zbar)&\equiv& R \left(U\: \phi(z,\zbar)\:U^{-1}\right),\qquad 
\cH'(\phi(z,\zbar)) \equiv R \left(U\: \cH(\phi(z,\zbar)) 
\: U^{-1}\right), \\ \nonumber
|0\rangle' &\equiv& U|0\rangle .
\end{eqnarray}  
Transformation by $U$ also leaves invariant the radial-ordered
commutator of two operators
\begin{eqnarray}
R\left(U [\cO_{1}(z),\cO_{2}(w)]U^{-1}\right) &=&
R\left(U\cO_{1}(z)U^{-1}U\cO_{2}(w)U^{-1}-
U\cO_{2}(w)U^{-1}U\cO_{1}(z)U^{-1}\right) \\
\nonumber 
&\equiv& R\left([\cO'_{1}(z),\cO'_{2}(w)]\right) .
\end{eqnarray}
Once this transformation is performed the result may then be analytically
continued back to Minkowski space.
We will give an explicit example of such a transformation for Abelian
bosonization in the following section.

We define two quantum field theories to be equivalent if they are 
related by a quantum canonical transformation and a finite
wavefunction renormalization.  Thus if the renormalized field
$\Phi'$ is defined as $\Phi'\equiv Z^{-\half}\phi'$ then
Green's functions for the free fields in the two theories,
in Minkowski space, are related by 
\beq
\label{free_Green_fn}
\langle 0 |' T \left[\Phi'(\vec{x}_{1},t_{1})\cdots
  \Phi'(\vec{x}_{n},t_{n})\right]
| 0 \rangle' = Z^{-{n\over 2}} \langle 0 | T \left[
\phi(\vec{x}_{1},t_{1})
\cdots \phi(\vec{x}_{n},t_{n}) \right]
| 0 \rangle .
\eeq
Furthermore the Green's functions for the interacting fields in the
new theory may, in principle, be calculated using 
Eqn. \eqref{free_Green_fn} and
Eqn. \eqref{interact_Green} with the interaction Hamiltonian $H'_{int}$.

\section{Free fermion and boson fields}
We begin with a $1+1$ dimensional field theory containing both
one complex fermion field and one real boson field periodic as 
$x_{1} \rightarrow x_{1}+\sqrt{2}L$. 
The Hilbert space of this theory 
is a tensor product of the respective Fock spaces, with the vacuum
state 
\beq
| 0 \rangle =
| 0 \rangle_{\mathrm{boson}}\otimes| 0 \rangle_{\mathrm{fermion}}.
\eeq
In terms of light-cone coordinates the mode expansion for the 
single real scalar field
$\phi$ is
\beq
\phi(\xplus,\xminus)= q + \qbar + {1\over 2L}(p x^{+} + \pbar x^{-}) 
+ \phi_{0}(\xplus,\xminus)
\eeq
with 
\begin{eqnarray}
\phi_{0}(\xplus,\xminus)&\equiv&
{\imath\over
{2\sqrt{\pi}}}
\sum_{n\neq 0}{1\over n} \left[ \alpha_{n}\emath^{-{2\pi\imath\over
    L}n\xplus}
+ \albar_{n} \emath^{-{2\pi\imath\over
    L}n\xminus}\right] \\ \nonumber
&\equiv&\phi_{0}(\xplus)
+\phibar_{0}(\xminus).
\end{eqnarray}
We have split the field $\phi$ into its zero-modes, the axial charges
$q$ and $\qbar$ and their conjugate momenta $p$ and $\pbar$, and the 
remainder $\phi_{0}$.
Hermiticity of $\phi_{0}(x)$ implies $\alpha^{\dagger}_{n}=\alpha_{-n}$.

The non-vanishing commutation relations $[\alpha_{m},\alpha_{n}]
=m\delta_{m,-n}$, $[\albar_{m},\albar_{n}]
=m\delta_{m,-n}$, and $[q,p]=[\qbar ,\pbar ]=\imath$ 
gives the usual equal-time commutation relation
\beq
\left.[\phi(x^{0},x^{1}),\partial_{0}\phi(y^{0},y^{1})]\right|_{x^{0}=y^{0}}=
\imath\delta(x^{1}-y^{1}) .
\eeq

A complex fermion field may be expressed in terms of two real fields as 
$\psi(x)\equiv\psi^{1}(x)+\imath\psi^{2}(x)$.  The mode expansion for
the real fermion fields $\psi^{i}(x)$, $i=1,2$ is 
\beq
\psi^{i}(\xplus,\xminus)={1\over 2^{3\over 4}}\left({1\over
    L}\right)^{\half}\sum_{n}\left[ \left(\begin{array}{c}
1 \\ 0 \end{array}\right)b_{n}^{i}
\emath^{-{2\pi\imath\over L}n\xplus} + \left(\begin{array}{c} 0 \\ 1
  \end{array}\right)\bbar_{n}^{i}\emath^{-{2\pi\imath\over L}n\xminus}
\right] .
\eeq
Again, since $\psi^{i}(x)$ are hermitian
$(b^{i}_{n})^{\dagger}=b^{i}_{-n}$ and the sum extends over integer 
values for Ramond (R) fields and half-integer values for Neveu-Schwarz
(NS) fields.
The non-vanishing anticommutators for the fermion creation and
annihilation operators are
$\{b^{i}_{m},b^{j}_{n}\}=\delta^{ij}\delta_{m,-n}$ and likewise for
$\bbar^{i}_{m}$.  The equal-time anticommutator for the real fermion fields is
\beq 
\left.\{\psi^{i}_{\alpha}(x^{0},x^{1}),\psi^{j}_{\beta}(y^{0},y^{1})\}
\right|_{x^{0}=y^{0}}= \half\delta^{ij}\delta_{\alpha\beta}\delta(x^{1}-y^{1})
\eeq
which gives for the complex field
\beq
\left.\{\psi_{\alpha}(x^{0},x^{1}),\psi_{\beta}^{\dagger}(y^{0},y^{1})\}
\right|_{x^{0}=y^{0}}=\delta_{\alpha\beta}\delta(x^{1}-y^{1}).
\eeq

The light-cone fermion current is defined as \footnote{Repeated
  indices imply summation over them throughout this paper.  The
  following convention for the gamma matrices is used:
  $\gamma^{0}\equiv\sigma_{x}$, $\gamma^{1}\equiv-\imath\sigma_{y}$, 
$\gamma^{5}\equiv\sigma_{z}$.}
\beq
\label{fermion_current}
J_{f}(\xplus)\equiv\imath\sqrt{2}L:\psi^{i}_{L}(\xplus)T_{ij}\psi^{j}
_{L}(\xplus):
,\qquad T \equiv \left(\begin{array}{cc}0&1\\-1&0\end{array}\right)
\eeq
with $\psi_{L,R}\equiv\psi_{1,2}$.
This current generates $U_{L}(1)\sim SO_{L}(2)$ symmetry
transformations $\psi_{L} \rightarrow \emath^{\imath\alpha}\psi_{L}$
and likewise $\Jbar_{f}(\xminus)$ generates $SO_{R}(2)$ transformations of the
fermion field.  Since $J_{f}$ and $\Jbar_{f}$ satisfy the wave equation $\Box
J_{f}=\Box \Jbar_{f}=0$ they may each be divided into positive and 
negative frequency modes and a normal-ordering defined for modes of
each current separately. 

We also define a boson current 
\beq
\label{boson_current}
J_{b}(\xplus)\equiv {L\over\sqrt{\pi}}\partial_{+}\phi(\xplus)
\eeq
that generates a shift in the boson field by a constant 
$\epsilon$, $\phi(\xplus)\rightarrow\phi(\xplus)+\epsilon$.  

Next, we will use the correspondence between a field theory 
of massless real bosons or fermions in Minkowski spacetime and 
conformal field theory in order to simplify calculations. 
After a Wick rotation, $x^{0}_{E}=-\imath x^{0}$ and defining 
$z\equiv\exp\left({2\pi\imath\over L}\xplus_{E}\right)$ we
have
\beq
\psi^{i}_{L}(\xplus)\rightarrow\psi^{i}(z) = {1\over 2^{3\over 4}}
\left({1\over
    L}\right)^{\half}\sum_{n}b^{i}_{n}z^{-n} 
\eeq
and similarily for $\psi^{i}_{R}(\xminus)$.
The Wick contraction for these fermion fields is defined as
\beq
\label{fermion_contraction}
|z|>|w|,\qquad \psi^{i}(z)\psi^{j}(w) = :\psi^{i}(z)\psi^{j}(w): + 
\delta^{ij}\Delta(z,w).
\eeq
For R-fields
\beq
\Delta(z,w)=\left({1\over{2^{5\over 2}L}}\right){z+w\over{z-w}}
\eeq
and for NS fields
\beq
\Delta(z,w)=\left({1\over{2^{3\over2}}L}\right){\sqrt{zw}\over{z-w}}.
\eeq

Likewise, after the change of variables,
$(\xplus,\xminus)\rightarrow(z,\zbar)$  
\beq
\phi_{0}(\xplus)\rightarrow \phi_{0}(z) = {\imath\over{2\sqrt{\pi}}}
\sum_{n\neq 0} {1\over n} \alpha_{n} z^{-n}. 
\eeq
The Wick contraction for $\phi_{0}(z)$ is
\beq
\label{boson_contraction}
|z|>|w|,\qquad \phi_{0}(z)\phi_{0}(w) = :\phi_{0}(z)\phi_{0}(w): 
-{1\over{4\pi}}
\ln\left(1-{w\over z}\right) . 
\eeq

The operator $U(\beta)$ which implements the quantum canonical
transformation in the conformal field theory is defined as the
radial-ordered exponential 
\beq
U(\beta)\equiv\mathrm{R}\exp\left[\imath2\sqrt{\pi}\beta
\left(\oint_{C} {dz\over
{2\pi\imath z}}
\phi_{0}(z)J_{f}(z) - \oint_{C} {d\zbar\over{2\pi\imath
    \zbar}}\phibar_{0}(\zbar)\Jbar_{f}
(\zbar)\right)\right] . 
\eeq
The contour C is chosen to be such that the complex coordinates of all 
other operators lie within it, {\it i.e.}, the limit $x^{0}_{min.}
\rightarrow \infty$ is taken.   $U(\beta)$ is unitary and its inverse 
results from $U(\beta)$ by taking $\imath
\rightarrow -\imath$ and integrating over a contour $C$ in which the
coordinates of the other operators lie outside of it, {\it i.e.}, in 
the limit $x^{0}_{max.} \rightarrow -\infty$.

One of the main properties of $U(\beta)$ is that it transforms the
fermion current $J_{f}$ 
and the boson current $J_{b}$ into one another.  This may be seen by 
evaluating Eqn. \eqref{operator_trans} for $U(\beta)$.  Using the definition 
of the currents
Eqns. \eqref{fermion_current} and \eqref{boson_current}, and
the Wick contractions Eqns. \eqref{fermion_contraction} and
\eqref{boson_contraction}, we find the following transformations of the
currents:
\begin{eqnarray}
U(\beta)\left(\begin{array}{c}J_{f}(z) \\ J_{b}(z)
  \end{array}
\right) U^{-1}(\beta) 
&=& \left(\begin{array}{cc} \cos\beta & \sin\beta \\ 
-\sin\beta & \cos\beta \end{array}\right) 
\left(\begin{array}{c}J_{f}(z) \\ J_{b}(z)
  \end{array}\right), \\ \nonumber
U(\beta)\left(\begin{array}{c}\Jbar_{f}(\zbar) \\ \Jbar_{b}(\zbar)
  \end{array}
\right) U^{-1}(\beta) 
&=& \left(\begin{array}{cc} \cos\beta & -\sin\beta \\ 
\sin\beta & \cos\beta \end{array}\right) 
\left(\begin{array}{c}\Jbar_{f}(z) \\ \Jbar_{b}(z)
  \end{array}\right) .
\end{eqnarray}
As expected, this transformation preserves the commutators of the
boson and fermion currents
\beq
[J_{f}(z),J_{f}(w)]= [J_{b}(z),J_{b}(w)] = {zw\over{(z-w)^{2}}},
\qquad [J_{f}(z),J_{b}(w)] = 0
\eeq
and likewise for the antiholomorphic currents.

Starting from a theory with only a free, massless complex fermion, we examine
how the Hamiltonian density transforms.  
In general the Hamiltonian density in terms of the light-cone components of the
energy-momentum tensor $\theta_{\mu\nu}$ is
\beq
\cH=\half\left(\theta_{++}(\xplus)+\theta_{--}(\xminus)\right) .
\eeq
The Hamiltonian density for the fermion field is 
\beq
\cH_{0}^{f} = {\imath\over 2}\left(\psi^{\dagger}\partial_{0}\psi 
- (\partial_{0}\psi^{\dagger})\psi\right)
\eeq 
which implies that $\theta_{\pm\pm}=\imath\sqrt{2}
:\psi^{i}_{\pm}\partial_{\pm}\psi^{i}_{\pm}:$ in terms of the real
fermion fields. 
The fermion Hamiltonian density may then be shown to be equal to the
Sugawara form
in terms of the currents of Eqn. \eqref{fermion_current} 
\cite{Sugawara_form,Goddard_Olive}
\beq
\cH_{0}^{f} = {\pi\over{2L^{2}}}\left(\ddagger J_{f}^{2}(\xplus)\ddagger 
+ \ddagger \Jbar_{f}^{2}(\xminus)\ddagger\right).
\eeq
The normal-ordering $\ddagger$ is with respect to modes of
$J(x^{+})$ and $\Jbar(x^{-})$, {\it i.e.}, with $J(\xplus)\equiv\sum_{n}J_{n}
\emath^{-{{2\pi\imath}\over L}\xplus}$ the $J_{n}$ with positive $n$
are placed to the right.
Choosing $\beta={\pi\over2}$ the transformed fermion Hamiltonian
density becomes
\begin{eqnarray}
U({\pi\over 2})\cH_{0}^{f}U^{-1}({\pi\over 2}) &=& 
{\pi\over{2L^{2}}}\left(\ddagger
J_{b}^{2}(\xplus)\ddagger 
+ \ddagger \Jbar_{b}^{2}(x_{-})\ddagger\right) \\ \nonumber
&=& \half\left(:(\partial_{+}\phi(\xplus))^{2}: 
+ :(\partial_{-}\phibar(\xminus))^{2}:\right) \\ \nonumber
&=& \half
:\partial_{\mu}\phi(x^{0},x^{1})
\partial^{\mu}\phi(x^{0},x^{1}): \\ \nonumber
&\equiv& \cH_{0}^{b} . 
\end{eqnarray}
The second line follows from the definition of the boson current,
Eqn. \eqref{boson_current}, and the fact that the usual normal-ordering 
with respect to
modes of $\phi$ is identical to the normal-ordering 
with respect to modes of $J_{b}$.  Thus the quantum canonical
transformation $U({\pi\over2})$ transforms the Hamiltonian for a free
complex fermion field into one for a free real boson field.
The fermion field decouples, {\it i.e.}, has no dynamics, and the
theory is equivalent to that for a free boson field.

\section{Complete Bosonization of the Massive Thirring model}
We next examine how the massive Thirring model transforms 
under $U({\pi\over 2})$, or complete bosonization.
The Hamiltonian density for the massive Thirring model is 
\begin{eqnarray}
\cH_{\mathrm{Thirring}}&=&\cH_{0}^{f}+m:\psibar\psi:+\half g
:\psibar\gamma_{\mu}\psi\psibar\gamma^{\mu}\psi: \\ \nonumber
&=&
\cH_{0}^{f}+\imath m \left(\psi_{L}^{i}T_{ij}\psi_{R}^{j} 
+ \psi_{R}^{i}T_{ij}\psi_{L}^{j}\right) 
-2g\left(:\psi_{L}^{i}T_{ij}\psi_{L}^{j}:\right)\left(:\psi_{R}^{i}
T_{ij}\psi_{R}^{j}:\right)        
\end{eqnarray}
where the Hamiltonian density is expressed in terms of the real chiral fermion 
fields.  
As discussed in Ref. \cite{Coleman}, the coupling $g$ must be greater
than $-\pi/2$ for the Thirring model with nonzero mass in order for the 
theory to be well defined. 
Both the mass and current-current terms have been regularized by
normal-ordering.  We also use the interaction representation with
massless fields in which these terms are treated as interactions.

First considering the massless case, the Hamiltonian density is  
\begin{eqnarray}
\label{massless_Th_H}
\cH_{m=0\ \mathrm{ Thirring}}&=&\cH_{0}^{f}-2g
\left(\psi_{L}^{i}T_{ij}\psi_{L}^{j}\right)
\left(\psi_{R}^{i}T_{ij}\psi_{R}^{j}\right) \\ \nonumber
&=&{\pi\over{2L^{2}}}\left(\ddagger J_{f}^{2}(\xplus)\ddagger 
+ \ddagger \Jbar_{f}^{2}(\xminus)\ddagger
+{2g\over\pi}J_{f}(\xplus)\Jbar_{f}(\xminus)\right).
\end{eqnarray}
Transforming by $U({\pi\over2})$ results in  
\begin{eqnarray}
\label{trans_massless_Th_H}
\cH'_{m=0\ \mathrm{ Thirring}} &=& U({\pi\over2})\cH_{m=0\ \mathrm
  { Thirring}}
U^{-1}({\pi\over2}) \\ \nonumber 
&=& 
{\pi\over{2L^{2}}}\left(\ddagger J_{b}^{2}(\xplus)\ddagger 
+ \ddagger \Jbar_{b}^{2}(\xminus)\ddagger
-{2g\over\pi}\ddagger J_{b}(\xplus)
\Jbar_{b}(\xminus)\ddagger \right) \\ \nonumber
&=& \half\left(:(\partial_{+}\phi)^{2}: +
  :(\partial_{-}\phi)^{2}:\right) 
-{g\over\pi} :\partial_{+}\phi\partial_{-}\phi: .
\end{eqnarray}
The corresponding Lagrangian density is 
\beq
\cL'_{m=0\ \mathrm{ Thirring}}=\left(\half + {g\over{2\pi}}\right)
:\partial_{\mu}\phi\partial^{\mu}\phi:,  
\eeq
which after a finite wavefunction renormalization becomes the
Lagrangian density of a free massless boson field.

In order to transform the mass term we use the methods of
Ref. \cite{Thirring_Currents} to express products of the fermion
fields in terms of the currents.  We next repeat the derivation 
for our case: a complex fermion field on a compact space and
currents defined by normal-ordered products.   
$J_{f}$ may be divided into the positive frequency 
modes $J_{f}^{(+)}$ 
and negative
frequency modes $J_{f}^{(-)}$, containing, respectively, creation and
annihilation operators.  The zero mode of $J_{f}$ is divided evenly
between $J_{f}^{(+)}$ and $J_{f}^{(-)}$.  The normal-ordered product
of a current and 
a fermion field is defined as 
\beq
\ddagger J_{f}(z)\psi^{i}(z)\ddagger \equiv J_{f}^{(+)}(z)
\psi^{i}(z) 
+ \psi^{i}(z)J_{f}^{(-)}(z).
\eeq
Using the Sugawara correspondence $\ddagger J_{f}(z)J_{f}(z) \ddagger  =
2^{3\over 2}L :z\left(\partial_{z} \psi^{i}(z)\right)
\psi^{i}(z):+\epsilon$, with $\epsilon=0,{1\over 4}$ for, respectively, NS 
or R fields, and the above definition of normal-ordering gives
\begin{eqnarray}
\half\left[\oint {dz\over {2\pi\imath z}} \ddagger J_{f}^{2}(z) \ddagger,
\psi^{i}_{L}(w)\right] &=& -\imath T^{ij}\ddagger J_{f}(w)
\psi^{j}_{L}(w)\ddagger \\ \nonumber
&=& w\partial_{w}\psi^{i}_{L}(w).
\end{eqnarray}
Integrating this equation and expressing the result in terms of the
original complex fermion field $\psi(z)=\psi^{1}(z)+\imath\psi^{2}(z)$   
\beq
\psi_{L}(z)=\ddagger \exp\left(-\int_{z_{0}}^{z} {d\xi\over \xi}
J_{f}(\xi)\right)
\psi_{L}(z_{0})\ddagger.
\eeq
This suggests the following expression
\beq
\psi_{L}(z)\psi_{L}^{\dagger}(w) = F(z,w)
\ddagger\exp\left(-\int_{w}^{z}{d\xi\over \xi} 
J_{f}(\xi)
\right)\ddagger
\eeq
with 
\beq
F(z,w)=\exp\left(\int_{w}^{z}{d\xi\over \xi}
J^{(+)}_{f}(\xi)\right)\psi_{L}(z)\psi_{L}^{\dagger}(w)
\exp\left(\int_{w}^{z}{d\xi\over \xi}
J^{(-)}_{f}(\xi)\right).
\label{F_def}
\eeq
$F(z,w)$ is a function of the charges since it commutes 
with $J(z)$ 
and if one assumes, as 
in Ref. \cite{Thirring_Currents}, that any operator that commutes with
the currents is a function only of the charges.  
$F(z,w)$ also commutes
with $\psi(u)$ for $u\neq z,w$ so it is actually a c-number function.
Finally with the commutators 
\begin{eqnarray}
\left[J_{f}^{(+)}(z),\psi_{L}(w)\right] &=& 
\left({w\over {w-z}} - \half\right)\psi_{L}(w), \\ \nonumber
\left[J_{f}^{(-)}(z),\psi_{L}^{\dagger}(w)\right] &=& 
\left({z\over {z-w}} - \half\right)\psi_{L}^{\dagger}(w)
\end{eqnarray}
and the definition of $F(z,w)$ in Eqn. \eqref{F_def} one obtains the 
following differential equations
\begin{eqnarray}
\partial_{z}F(z,w) &=& \left(-{1\over{z-w}}+{1\over{2z}}\right)
F(z,w),\\ \nonumber
\partial_{w}F(z,w) &=& \left({1\over{z-w}}+{1\over{2w}}\right)
F(z,w)
\end{eqnarray}
with the solution
\beq
F(z,w)=f_{0}{\sqrt{zw}\over{z-w}},
\eeq
where $f_{0}$ is a constant.
Thus we arrive at 
\beq
\label{psi_soln}
\psi_{L}(z)\psi_{L}^{\dagger}(w) = f_{0}{\sqrt{zw}\over{z-w}}
\ddagger\exp\left(-\int_{w}^{z}{d\xi\over \xi}
J_{f}(\xi)\right)\ddagger .
\eeq
Although this solution appears to be
non-local it is in fact bi-local since it is independent of the
integration path from $w$ to $z$.
We next define $\sigma_{+}(x^{+},x^{-})\equiv
\psi_{L}^{\dagger}(x^{+})\psi_{R}(x^{-})$ and $\sigma_{-}(x^{+},x^{-})\equiv
\psi_{R}^{\dagger}(x^{-})\psi_{L}(x^{+})$.  Non-zero vacuum
expectation values of
products of these operators must contain an equal number of
$\sigma_{+}$ and $\sigma_{-}$ because of the $SO_{L}(2)\times
SO_{R}(2)$ symmetry.
We consider the product of operators  
\beq
\label{sigma_pm}
\sigma_{+}(z,\zbar)\sigma_{-}(w,\wbar) = f_{0}^{2}
\left|{\sqrt{zw}\over{z-w}}\right|^{2}\ddagger 
\exp\left(\int_{w}^{z}{d\xi\over \xi} J_{f}(\xi)\right)\ddagger
\ddagger 
\exp\left(-\int_{\wbar}^{\zbar}{d\xibar\over \xibar} 
\Jbar_{f}(\xibar)\right)\ddagger
\eeq
with the equality following from the 
solution of Eqn. \eqref{psi_soln}.
The value of $f_{0}$ may be found by
comparing the vacuum expectation value of this operator   
\begin{eqnarray}
\langle 0 | \sigma_{+}(z,\zbar)\sigma_{-}(w,\wbar) | 0 \rangle &=& 
\langle 0 | \psi^{i}(z) \psi^{i}(w) | 0 \rangle 
\langle 0 | \psibar^{i}(\zbar)\psibar^{i}(\wbar) | 0 \rangle \\ \nonumber
&=& {1\over {2 L^{2}}} \left| {\sqrt{zw}\over{z-w}} \right|^{2}
\end{eqnarray} 
with the operator in terms of the currents in Eqn. \eqref{sigma_pm}.
Because the vacuum expectation value of normal-ordered products of current
operators vanishes, $f_{0} = 1/(\sqrt{2}L)$. 

We next find the transformation of the operator in Eqn. \eqref{sigma_pm}
by $U({\pi\over 2})$ to get the corresponding operator in the bosonic
theory 
\begin{eqnarray}
\label{sigma_trans}
&U&({\pi\over 2})\sigma_{+}(z,\zbar)\sigma_{-}(w,\wbar)U^{-1}({\pi\over
  2})\\ \nonumber
&=& {1\over{2L^{2}}}\left|{\sqrt{zw}\over{z-w}}\right|^{2} 
:\exp\left(2\sqrt{\pi}\imath(\phi_{0}(z)-\phi_{0}(w))\right): 
:\exp\left(2\sqrt{\pi}\imath(\phibar_{0}(\zbar)
-\phibar_{0}(\wbar))\right):\\ \nonumber
&=& {1\over{2L^{2}}}\left|\sqrt{{w\over z}}\right|^{2} 
:\exp\left(2\sqrt{\pi}\imath(\phi_{0}(z)+\phibar_{0}(\zbar))\right): 
:\exp\left(-2\sqrt{\pi}\imath(\phi_{0}(w)
+\phibar_{0}(\wbar))\right): .
\end{eqnarray}
The last equality may be derived using the relation
\begin{eqnarray}
:\emath^{2\sqrt{\pi}\imath\phi_{0}(z)}: 
:\emath^{-2\sqrt{\pi}\imath\phi_{0}(w)}: &=& \emath^{2\sqrt{\pi}
\imath\psi_{0}^{(+)}(z)} \emath^{2\sqrt{\pi}\imath\psi_{0}^{(-)}(z)} 
\emath^{-2\sqrt{\pi}\imath\psi_{0}^{(+)}(w)} 
\emath^{-2\sqrt{\pi}\imath\psi_{0}^{(-)}(w)} \\ \nonumber
&=& \emath^{4\pi[\phi_{0}^{(-)}(z),\phi_{0}^{(+)}(w)]} 
:\emath^{2\sqrt{\pi}\imath(\phi_{0}(z)-\phi_{0}(w))}: \\ \nonumber
&=& \left({z\over{z-w}}\right):\emath^{2\sqrt{\pi}\imath
(\phi_{0}(z)-\phi_{0}(w))}: . 
\end{eqnarray}
$\phi_{0}^{(+)}$ and $\phi_{0}^{(-)}$ are, respectively, the positive
and negative frequency modes of $\phi_{0}$.

Since Eqn. \eqref{sigma_trans} is valid for arbitrary $|z| > |w|$ the
transformation of the operators $\sigma_{+}$ and $\sigma_{-}$ are 
\begin{eqnarray}
U({\pi\over 2})\sigma_{+}(z,\zbar)U^{-1}({\pi\over 2}) 
&=& {1\over {\sqrt{2}L}}(\sqrt{z\zbar})^{-1}
:\exp\left(2\sqrt{\pi}\imath(\phi_{0}(z)
+\phibar_{0}(\zbar))\right):, \\ \nonumber
U({\pi\over 2})\sigma_{-}(z,\zbar)U^{-1}({\pi\over 2}) 
&=& {1\over {\sqrt{2}L}}\sqrt{z\zbar}:\exp\left(-2\sqrt{\pi}\imath(\phi_{0}(z)
+\phibar_{0}(\zbar))\right): .
\end{eqnarray}

Finally, to compare with previous results in the literature we add 
the zero-modes to $\phi_{0}$ and $\phibar_{0}$
\beq
\phi(z)  \equiv q + p{1\over{4\pi\imath}}\ln z + \phi_{0}(z),\qquad
\phibar(\zbar) \equiv \qbar + \pbar {1\over{4\pi\imath}}\ln \zbar 
+ \phibar_{0}(\zbar )
\eeq
and define normal-ordered expressions to have $p$ to the right of
$q$.  Then the transformed operators become  
\begin{eqnarray}
U({\pi\over 2})\sigma_{+}(z,\zbar)U^{-1}({\pi\over 2}) 
&=& {1\over {\sqrt{2}L}}\sqrt{z\zbar}
:\exp\left(2\sqrt{\pi}\imath(\phi(z)
+\phibar(\zbar))\right):, \\ \nonumber
U({\pi\over 2})\sigma_{-}(z,\zbar)U^{-1}({\pi\over 2}) 
&=& {1\over {\sqrt{2}L}}\sqrt{z\zbar}:\exp\left(-2\sqrt{\pi}\imath(\phi(z)
+\phibar(\zbar))\right): .
\end{eqnarray}
This result agrees with the Frenkel-Kac fermionic vertex operator 
construction with fermion operators $\Psi$ in the transformed theory  
\cite{Frenkel,Goddard_Olive}
\begin{eqnarray}
\label{vertex_op}
\Psi_{L}(z) &=& 2^{-\fourth}L^{-\half}\sqrt{z}:\exp\left(-2\sqrt{\pi}
\imath\phi(z)\right):
\epsilon_{L}, \qquad 
\Psi_{L}^{\dagger}(z) = 2^{-\fourth}L^{-\half}\sqrt{z}:\exp\left(2\sqrt{\pi}
\imath\phi(z)\right):\epsilon_{L}, \\ \nonumber
\Psi_{R}(\zbar) &=&
2^{-\fourth}L^{-\half}\sqrt{\zbar}:\exp\left(2\sqrt{\pi}
\imath\phibar(\zbar)\right):
\epsilon_{R}, \qquad 
\Psi_{R}^{\dagger}(\zbar) = 2^{-\fourth}L^{-\half}\sqrt{\zbar}
:\exp\left(-2\sqrt{\pi}
\imath\phibar(\zbar)\right):\epsilon_{R} 
\end{eqnarray}
and $\epsilon_{L,R}$ are coordinate-independent fermionic
operators which satisfy $\{\epsilon_{L},\epsilon_{R}\}=0$ and 
$(\epsilon_{L})^{2}=(\epsilon_{R})^{2}=1$ \cite{Banks}.
The operators in Eqn. \eqref{vertex_op} then have the correct 
anticommutation relations for complex fermion fields.

All that remains to be done to obtain the complete bosonized
Hamiltonian density is a finite wavefunction renormalization of
$\phi(x)$.  The renormalized field $\Phi(x)$ is defined as $\Phi(x) = 
\alpha^{-1}\:\phi(x)$.  We use a method 
similar to that used by Coleman in order to regularize the expression 
$:\exp(2\sqrt{\pi}\imath\alpha\Phi(z)):$ whereby one uses 
Wick's theorem \cite{Coleman,Banks}
\beq
\exp\left(\imath\int J(x)\phi(x) d^{2}x\right) = :\exp\left(\imath
\int J(x)\phi(x) d^{2}x\right): \exp\left(-\half\int
J(x)\Delta(x-y;m)J(y)d^{2}x d^{2}y\right)
\eeq
with $J(x)$ set equal to a delta-function and the propagator
$\Delta(x;m)$, which is singular for spacelike $x^{2}\rightarrow 0$
replaced by the regulated propagator $\Delta^{R}(x;m,\Lambda)$, which
is finite in this limit
\beq  
\Delta^{R}(x;m,\Lambda) \equiv \Delta(x,m) - \Delta(x,m=\Lambda).
\eeq
$\Lambda$ is a large cutoff mass.  In our case $m=0$ and, with the 
propagator of
Eqn. \eqref{boson_contraction}, the result for the field $\phi_{0}$ is 
\beq
:\emath^{2\sqrt{\pi}\imath\phi_{0}(z,\zbar)}:_{\phi_{0}} = 
\left({L^{2}\Lambda^{2}\over 
      {4\pi^{2}}}\right)^{-\alpha^{2}+1}
:\emath^{2\sqrt{\pi}\imath\alpha\Phi_{0}(z,\zbar)}:_{\Phi_{0}}.
\eeq
The subscript on the normal-ordering symbol indicates with respect to
which field the expression is normal-ordered.  Adding in the
zero-modes one obtains
\beq
:\emath^{2\sqrt{\pi}\imath\phi(z,\zbar)}:_{\phi} = \left({L^{2}\Lambda^{2}
\over 
{4\pi^{2}}}\right)^{-\alpha^{2}+1}\left|z\right|^{\half(\alpha^{2}-1)}
:\emath^{2\sqrt{\pi}\imath\alpha\Phi(z,\zbar)}:_{\Phi}.
\eeq
After collecting the previous results with
$\alpha=\left(1+{g\over\pi}\right)^{-\half}$ and an analytical
continuation back to Minkowski space, the Lagrangian density
$\cL_{b}$ corresponding to the bosonized massive Thirring model is 
\beq
\label{full_boson_L}
\cL_{b} = \half:\partial_{\mu}\Phi\partial^{\mu}\Phi: 
-{\sqrt{2}m\over L} \left({L^{2}\Lambda^{2}\over{4\pi^{2}}}\right)
^{g\over{\pi+g}}\exp\left({\sqrt{2}\pi\over
    L}\left(1+{g\over\pi}\right)^{-1}x^{0}\right)
:\cos 2\sqrt{\pi}\left(1+{g\over\pi}\right)^{-\half}\Phi:_{\Phi} .
\eeq

The coefficient $a$ of $\cos a \Phi$ in the potential, which is 
renormalization scheme
independent, agrees with that derived using different 
methods \cite{Coleman,Mandel_Klaiber}.
The $x^{0}$ dependence of the unrenormalized potential term is
necessary for it to have a definite scaling dimension, {\it i.e.}, in
conformal field theory conformal weight $(\half,\half)$.  Wavefunction 
renormalization maintains this property however changes the scaling
dimension.   The operator in the potential term of 
Eqn. \eqref{full_boson_L} now has scaling dimension
$(1+{g\over\pi})$.  This term is therefore super-renormalizable only 
within the 
allowed range of the coupling, $g<{\pi\over 2}$. 
Finally, the finite wavefunction renormalization gives the
correctly normalized 
kinetic term in the Lagrangian, however the transformed fermion
operators in the new theory, $\Psi_{L,R}$, no longer satisfy fermionic
anticommutation relations.  
 
\section{Smooth bosonization of the Massive Thirring Model}
The previous solution of the massive Thirring model in terms of the 
currents may also be used to find a continuum of equivalent theories
which contain interactions between the boson and fermion field by
transforming the Thirring model Hamiltonian density by $U(\beta)$.  
The result for the massless Thirring model is   
\begin{eqnarray}
\label{trans_massless_beta}
U(\beta)\cH_{\mathrm{m=0\ Thirring}}U^{-1}(\beta) &=& \cos^{2}\beta 
\:\cH_{\mathrm{m=0\ Thirring}} + \sin^{2}\beta \:
\cH'_{\mathrm{m=0\ Thirring}} \\ \nonumber 
    &+& {1\over{2L^{2}}}\sin 2\beta \left[\pi(J_{f}J_{b} 
- \Jbar_{f}\Jbar_{b}) + g(J_{f}\Jbar_{b} - \Jbar_{f}J_{b})\right].
\end{eqnarray}
The Hamiltonian density for the massless Thirring model 
$\cH_{\mathrm{m=0\ Thirring}}$ and its bosonic equivalent 
$\cH'_{\mathrm{m=0\ Thirring}}$ are given in Eqns. \eqref{massless_Th_H} and 
\eqref{trans_massless_Th_H}, respectively.   One follows the same
procedure as for the case of complete bosonization to calculate the
transformed mass term, namely transform the fermion currents in the 
operator solution of 
$\sigma_{+}\sigma_{-}$, giving
\begin{eqnarray}
\label{sigma_beta_trans}
U(\beta)\sigma_{+}(z,\zbar)\sigma_{-}(w,\wbar)U^{-1}(\beta) &=& 
{1\over {2L^{2}}}\left|{\sqrt{zw}\over{z-w}}\right|^{2} 
\ddagger \exp\left(\int_{w}^{z}{d\xi\over\xi}\left(\cos\beta J_{f}(\xi) 
+\sin\beta J_{b}(\xi)\right)\right)\ddagger  \\ \nonumber
&\times&\ddagger \exp \left(\int_{\wbar}^{\zbar}{d\xibar\over\xibar}
\left(-\cos\beta \Jbar_{f}(\xibar) 
+\sin\beta \Jbar_{b}(\xibar)\right)\right)\ddagger .
\end{eqnarray}
Substituting the definition of the boson current
Eqn. \eqref{boson_current} and reordering the boson terms gives 
\begin{eqnarray}
\label{sigma_transform}
U(\beta)\sigma_{+}(z,\zbar)\sigma_{-}(w,\wbar)U^{-1}(\beta) &=& 
{1\over
  {2L^{2}}}|w||z|^{1-2\sin^{2}\beta}
\left|{1\over{z-w}}\right|^{2(1-\sin^{2}\beta)} \\ \nonumber
&\times&:\exp\left(2\sqrt{\pi}\imath\sin\beta(\phi_{0}(z)
+\phibar_{0}(\zbar))\right): \\ \nonumber
&\times&:\exp\left(-2\sqrt{\pi}\imath\sin\beta(\phi_{0}(w)
+\phibar_{0}(\wbar))\right): \\ \nonumber
&\times& \ddagger \exp\left(\cos\beta\int_{w}^{z}{d\xi\over
    \xi}J_{f}(\xi)\right)\ddagger 
\ddagger \exp\left(-\cos\beta\int_{\wbar}^{\zbar}
{d\xibar\over \xibar}\Jbar_{f}(\xibar)\right)\ddagger .
\end{eqnarray}
There is no simple operator expression for the normal-ordered exponents 
of the fermion currents in this equation in terms of the 
fermion fields $\psi_{L,R}$.  Therefore the fermionic sector of the 
transformed theory will be formulated in terms of $J_{f}$ and $\Jbar_{f}$ 
rather than $\psi_{L}$ and $\psi_{R}$.  This is not such a radial 
reformulation of the theory since operators with non-vanishing vacuum 
expectation values in the original theory, the massive Thirring model, may 
also be expressed entirely 
in terms of the fermion currents, using the equations given
previously.  Next, in order to obtain $\sigma_{\pm}$ in the new theory 
we separate the normal-ordered exponent of the fermion current, for
$|z|>|w|$ as  
\begin{eqnarray}
\ddagger\exp\left[\left(\cos\beta\int_{w}^{z}{d\xi\over\xi}J_{f}(\xi)
\right)\right]\ddagger &=&\ddagger\exp\cos\beta\left(\int_{\infty}^{z}
{d\xi\over\xi}J_{f}(\xi)-\int_{\infty}^{w}
{d\xi\over\xi}J_{f}(\xi)\right)\ddagger \\ \nonumber
 &=&\left({z\over{z-w}}\right)^{-\cos^{2}\beta}\ddagger\exp\left(\cos\beta
\int_{\infty}^{z}{d\xi\over\xi}J_{f}(\xi)\right)\ddagger \\ \nonumber
 &\times& \ddagger\exp\left(\cos\beta\int_{\infty}^{w}{d\xi\over\xi}
J_{f}(\xi)\right)\ddagger .
\end{eqnarray}
An adiabatic cutoff, $\emath^{-\epsilon\xi}$ with $\epsilon$ small, is 
implied in the integration.  After substituting this relation in
Eqn. \eqref{sigma_transform} the result is a product of two factors,
each with dependence on only one coordinate.  Therefore the
transformed $\sigma_{\pm}$ are
\begin{eqnarray}
\label{trans_sigma_unren}
U(\beta)\sigma_{+}(z,\zbar)U^{-1}(\beta) &=& {1\over {\sqrt{2}L}}|z|^{-1}
:\exp\left(2\sqrt{\pi}\imath\sin\beta(\phi_{0}(z)+\phibar_{0}(\zbar))\right):
\\ \nonumber
&\times&\ddagger\exp\left[\cos\beta\left(\int_{\infty}^{z}
{d\xi\over\xi}J_{f}(\xi)
-\int_{\infty}^{\zbar}{d\xibar\over\xibar}\Jbar_{f}(\xibar)
\right)\right], \ddagger \\ \nonumber
U(\beta)\sigma_{-}(z,\zbar)U^{-1}(\beta) &=& {1\over {\sqrt{2}L}}|z|
:\exp\left(-2\sqrt{\pi}\imath\sin\beta(\phi_{0}(z)+\phibar_{0}(\zbar))\right): 
\\ \nonumber
&\times&\ddagger\exp\left[\cos\beta\left(-\int_{\infty}^{z}{d\xi\over\xi}
J_{f}(\xi)+\int_{\infty}^{\zbar}{d\xibar\over\xibar}\Jbar_{f}(\xibar)
\right)\right] \ddagger  .\\ \nonumber 
\end{eqnarray}
This implies that the transformed fermion operators, for example $\Psi_{L}(z)=
U(\beta)\psi_{L}(z)U^{-1}(\beta)$, are 
\begin{eqnarray}
\Psi_{L}(z) &=&
2^{-\fourth}L^{-\half}\sqrt{z}:\exp\left(-2\sqrt{\pi}\imath
\sin\beta\phi_{0}(z)\right): \ddagger\exp\left(-\cos\beta\int_{\infty}^{z}
{d\xi\over\xi}J_{f}(\xi)\right)\ddagger\epsilon_{L}, \\ \nonumber
\Psi_{L}^{\dagger}(z) &=&
2^{-\fourth}L^{-\half}\sqrt{z}^{-1}:\exp\left(2\sqrt{\pi}\imath
\sin\beta\phi_{0}(z)\right): \ddagger\exp\left(\cos\beta\int_{\infty}^{z}
{d\xi\over\xi}J_{f}(\xi)\right)\ddagger \epsilon_{L}, \\ \nonumber
\Psi_{R}(\zbar) &=&
2^{-\fourth}L^{-\half}\sqrt{\zbar}^{-1}:\exp\left(2\sqrt{\pi}\imath
\sin\beta\phibar_{0}(\zbar)\right): \ddagger\exp\left(-\cos\beta
\int_{\infty}^{\zbar}
{d\xibar\over\xibar}\Jbar_{f}(\xibar)\right)\ddagger\epsilon_{R}, \\ \nonumber
\Psi_{R}^{\dagger}(\zbar) &=&
2^{-\fourth}L^{-\half}\sqrt{\zbar}:\exp\left(-2\sqrt{\pi}\imath
\sin\beta\phibar_{0}(\zbar)\right): \ddagger\exp\left(\cos\beta
\int_{\infty}^{\zbar}
{d\xibar\over\xibar}\Jbar_{f}(\xibar)\right)\ddagger \epsilon_{R} .
\end{eqnarray}
$\epsilon_{L}$ and $\epsilon_{R}$ have the same definition as in Eqn.
\eqref{vertex_op}.  These transformed fermion operators also satisfy 
the correct
anticommutation relations for complex fermion fields.

Next we perform a finite renormalization of both the fermion and
the boson currents.  The renormalization of the boson currents is
equivalent to renormalization of the boson field $\phi(z,\zbar)$.  The 
renormalized fields are defined as 
\beq 
J_{f}^{R}(z)=J_{f}(z)\cos\beta, \qquad
\Jbar_{f}^{R}(\zbar)=\Jbar_{f}(\zbar)
\cos\beta, \qquad
\Phi(z,\zbar)=\phi(z,\zbar)\sin\beta .
\eeq
Using the same method as in the last section to express the
normal-ordered exponentials of expressions containing $\phi$, $J_{f}$, 
and $\Jbar_{f}$ in terms of the renormalized quantities, Eqn. 
\eqref{trans_sigma_unren} becomes 
\begin{eqnarray}
\label{renorm_trans_sigma}
\left(U(\beta)\sigma_{+}(z,\zbar)U^{-1}(\beta)\right)_{\mathrm{ren.}}
&=& {1\over{\sqrt{2}L}}|z|^{-1}\left({{4\pi^{2}}\over{L^{2}
\Lambda^{2}}}\right):\exp\left(2\sqrt{\pi}\imath(\phi_{0}(z)
+\phibar_{0}(\zbar))\right):
\\ \nonumber
&\times&\ddagger\exp\left[\left(\int_{\infty}^{z}
{d\xi\over\xi}J_{f}(\xi)
-\int_{\infty}^{\zbar}{d\xibar\over\xibar}\Jbar_{f}(\xibar)
\right)\right] \ddagger, \\ \nonumber
\left(U(\beta)\sigma_{-}(z,\zbar)U^{-1}(\beta)\right)_{\mathrm{ren.}} 
&=& {1\over {\sqrt{2}L}}|z|\left({{4\pi^{2}}\over{L^{2}\Lambda^{2}}}\right)
:\exp\left(-2\sqrt{\pi}\imath(\phi_{0}(z)+\phibar_{0}(\zbar))\right): 
\\ \nonumber
&\times&\ddagger\exp\left[\left(-\int_{\infty}^{z}{d\xi\over\xi}
J_{f}(\xi)+\int_{\infty}^{\zbar}{d\xibar\over\xibar}\Jbar_{f}(\xibar)
\right)\right] \ddagger  .\\ \nonumber 
\end{eqnarray}
Collecting the transformed operators of 
Eqns. \eqref{trans_massless_beta} and \eqref{renorm_trans_sigma} in
terms of the renormalized fields the transformed Hamiltonian density
$\cH'_{\mathrm{ren.}}$ in Minkowski space becomes  
\begin{eqnarray}
\label{beta_trans_H}
\cH'_{\mathrm{ren.}} &=& \half \left(:(\partial_{+}\Phi)^{2}: + 
:(\partial_{-}\Phi)^{2}:\right) -{g\over\pi}:\partial_{+}\Phi
\partial_{-}\Phi: \\ \nonumber 
&+& {\pi\over{2L^{2}}}\left(\ddagger (J_{f}^{R})^{2}\ddagger
  + \ddagger (\Jbar_{f}^{R})^{2}\ddagger +
  {{2g}\over\pi}J_{f}^{R}\Jbar_{f}^{R}\right) \\ \nonumber
&+& {\sqrt{\pi}\over
  L}\left[J_{f}^{R}(\partial_{+}\Phi+{g\over\pi}\partial_{-}\Phi) -
  \Jbar_{f}^{R}(\partial_{-}\Phi+{g\over\pi}\partial_{+}\Phi)\right]
\\ \nonumber
&+& {m\over{\sqrt{2}L}}\left({{4\pi^{2}}\over{L^{2}\Lambda^{2}}}\right)\left[
\exp\left(-{{2\sqrt{2}\pi}\over L}x^{0}\right) :\exp\left(2\sqrt{\pi}
\imath\Phi\right):\right. \\ \nonumber
&\times&\ddagger\exp{{2\pi\imath}\over
L}\left(\int_{\infty}^{x^{+}}d\xi^{+}J_{f}^{R}(\xi^{+})-
\int_{\infty}^{x^{-}}d\xi^{-}\Jbar_{f}^{R}(\xi^{-})
\right)\ddagger
\\ \nonumber
&+& \exp\left({{2\sqrt{2}\pi}\over L}x^{0}\right)
:\left.\exp\left(-2\sqrt{\pi}\imath\Phi\right): \ddagger\exp
{{2\pi\imath}\over
  L}\left(-\int_{\infty}^{x^{+}}d\xi^{+}J_{f}^{R}(\xi^{+}) +
  \int_{\infty}^{x^{-}}d\xi^{-}\Jbar_{f}^{R}(\xi^{-})\right)\ddagger\right] .
\end{eqnarray}

\section{Discussion}
If we consider parity transformations $\cP$ we find that 
$\cP U(\beta)\cP^{-1} = U(-\beta)$ for a scalar $\phi$.  However $U(\beta)$ 
is invariant under a parity transformation if $\phi$ is a
pseudoscalar, {\it i.e.}, $\cP \phi(x^{0},x^{1}) \cP^{-1} = 
-\phi(x^{0},-x^{1})$.  Therefore since we start with the parity
invariant massive Thirring model, the transformed
Hamiltonian density of Eqn. \eqref{beta_trans_H} is also parity
invariant, but only if $\phi$ is a pseudoscalar.  This is particularly
interesting for the suggestion of Damgaard {\it et al.} of using smooth
bosonization to construct exact Cheshire cat bag models in $1+1$
dimensions to describe QCD bound states \cite{Smooth_boson}.  In these
models the interior 
of the bag has fermion fields, namely quarks, whereas outside the
bag the relevant degrees of freedom are the scalar fields, namely
mesons.  Smooth bosonization allows a continuous transition from the
fermionic theory to the bosonic one with interactions between them
only near the bag surface.  This interpretation is consistent
with the method of smooth bosonization presented in this paper since
the meson field, $\phi$, has the correct parity.  

We have also presented an explicit unitary operator implementing this
bosonization transformation.  It would be interesting to investigate
similar operators for other duality transformations.  
Although the
quantum canonical transformation in this paper is linear (in the
currents), as are all such transformations in quantum field theory 
presented to date,
the method of transformation by a unitary operator opens up the
possibility of finding transformations not within this class.
Finally, it is important to understand the exact relation 
between the various
methods of demonstrating duality: path integral manipulations, linear
canonical transformations, and operator transformations.  This would
allow comparison between the results derived using these complementary 
methods. 

\section*{Acknowledgements}
We thank R. Sasaki and J. Ding for useful discussions and comments.
This work was supported by the National Science Foundation under 
grant no. 9415225 and the Japan Society for the Promotion of
Science.


\begin{thebibliography}{99}
\bibitem{NA_bosonization}E. Witten, Comm. Math. Phys. {\bf 92}, 455 (1984).
\bibitem{S_duality}E. Witten, ``On S Duality in Abelian Gauge
  Theory'', Princeton I.A.S. preprint IASSNS-HEP--95-36, hep-th/9505186.
\bibitem{path_int_boson}C.P. Burgess and F. Quevedo, Nucl. Phys. {\bf
    B421} 373 (1994); M.R. Garousi, Phys. Rev. {\bf D53} 2173 (1996).
\bibitem{T_duality}E. Alvarez, L. Alvarez-Gaum\'{e}, and Y. Lozano,
  Nucl. Phys. (Proc. Supp.) {\bf B41A} 1 (1995); A. Giveon,
  M. Porrati, and E. Rabinovici, Phys. Rep. {\bf 244} 77 (1994).
\bibitem{Anderson}A. Anderson, Phys. Lett. {\bf B305} 67 (1993);
  {\bf B319} 157 (1993)
\bibitem{Coleman}S. Coleman, Phys. Rev. {\bf D11}, 2088 (1975).
\bibitem{Mandel_Klaiber}S. Mandelstam, {\bf D11}, 3026 (1975);
  B. Klaiber, in {\it Lectures in Theoretical Physics}, proceedings of 
  the Tenth Boulder Summer Institute for Theoretical Physics, edited
  by A. Barut and W. Brittin, (Gordon and Breach, New York, 1968), Vol.X-A.
\bibitem{Banks}T. Banks, D. Horn, and H. Neuberger, Nucl. Phys. {\bf
    B108}, 119 (1976).
\bibitem{R_quantization}P.A.M. Dirac, Rev. Mod. Phys. {\bf 21}, 392
  (1949); 
S. Fubini, R. Jackiw, and A. Hanson, Phys. Rev. {\bf D7}, 1732 (1973); 
C. Lovelace, Nucl. Phys. {\bf B99}, 109 (1975).
\bibitem{Sugawara_form}
H. Sugawara, Phys. Rev. {\bf 170}, 1659 (1968);
C. Sommerfield, Phys. Rev. {\bf 176}, 2019 (1968).
\bibitem{Goddard_Olive} P. Goddard and D. Olive,
  Int. Journ. Mod. Phys., {\bf A1}, 303 (1986).
\bibitem{Thirring_Currents}G.F. Dell'Antonio, Y. Frishman and
  D. Zwanziger, Phys. Rev. {\bf D6}, 988 (1972).
\bibitem{Frenkel}I.B. Frenkel, J. Funct. Anal. {\bf 44}, 259 (1981).
\bibitem{Smooth_boson}P.H. Damgaard, H.B. Nielsen, and R. Sollacher,
  Nucl. Phys. {\bf B385}, 227 (1992);  P.H. Damgaard, H.B. Nielsen, 
and R. Sollacher, Phys. Lett. {\bf B296}, 132 (1992).
\end{thebibliography}
\end{document}